\documentclass[sigconf]{acmart}

\usepackage{balance}
\usepackage{booktabs}
\usepackage{multirow}
\usepackage{subcaption}
\usepackage[ruled]{algorithm2e}

\newlength\mylen
\newcommand\myinput[1]{%
	\settowidth\mylen{\KwIn{}}%
	\setlength\hangindent{\mylen}%
	\hspace*{\mylen}#1\\}
	
\def\BibTeX{{\rm B\kern-.05em{\sc i\kern-.025em b}\kern-.08emT\kern-.1667em\lower.7ex\hbox{E}\kern-.125emX}}
    
\copyrightyear{2019}
\acmYear{2019}
\setcopyright{acmcopyright}
\acmConference[CIKM '19]{The 28th ACM International Conference on Information and Knowledge Management}{November 3--7, 2019}{Beijing, China}
\acmBooktitle{The 28th ACM International Conference on Information and Knowledge Management (CIKM'19), November 3--7, 2019, Beijing, China}
\acmPrice{15.00}
\acmDOI{10.1145/3357384.3357893}
\acmISBN{978-1-4503-6976-3/19/11}

\begin{document}

\fancyhead{}

\title{BHIN2vec: Balancing the Type of Relation in Heterogeneous Information Network}

\author{Seonghyeon Lee}
\email{sh0416@postech.ac.kr}
\affiliation{%
    \institution{Pohang University of Science and Technology}
    \city{Pohang}
    \country{South Korea}
}
\author{Chanyoung Park}
\email{pcy1302@illinois.edu}
\affiliation{%
  \institution{University of Illinois at Urbana-Champaign}
  \city{Urbana}
  \country{USA}
}
\author{Hwanjo Yu}
\authornote{Corresponding author}
\email{hwanjoyu@postech.ac.kr}
\affiliation{%
	\institution{Pohang University of Science and Technology}
	\city{Pohang}
    \country{South Korea}
}

%
\renewcommand{\shortauthors}{Seonghyeon, et al.}

%
\begin{abstract}
The goal of network embedding is to transform nodes in a network to a low-dimensional embedding vectors. Recently, heterogeneous network has shown to be effective in representing diverse information in data. However, heterogeneous network embedding suffers from the imbalance issue, i.e. the size of relation types (or the number of edges in the network regarding the type) is imbalanced. In this paper, we devise a new heterogeneous network embedding method, called BHIN2vec, which considers the balance among all relation types in a network. We view the heterogeneous network embedding as simultaneously solving multiple tasks in which each task corresponds to each relation type in a network. After splitting the skip-gram loss into multiple losses corresponding to different tasks, we propose a novel random-walk strategy to focus on the tasks with high loss values by considering the relative training ratio. Unlike previous random walk strategies, our proposed random-walk strategy generates training samples according to the relative training ratio among different tasks, which results in a balanced training for the node embedding. Our extensive experiments on node classification and recommendation demonstrate the superiority of BHIN2vec compared to the state-of-the-art methods. Also, based on the relative training ratio, we analyze how much each relation type is represented in the embedding space. 
\end{abstract}

%
%

\ccsdesc[500]{Theory of computation~Unsupervised learning and clustering}
\ccsdesc[500]{Computing methodologies~Knowledge representation and reasoning}

%
\keywords{network embedding, heterogeneous network, random-walk strategy, multitask learning, inverse training ratio, stochastic matrix}

%
\maketitle

\begin{figure}
	\centering
	\includegraphics[width=\linewidth]{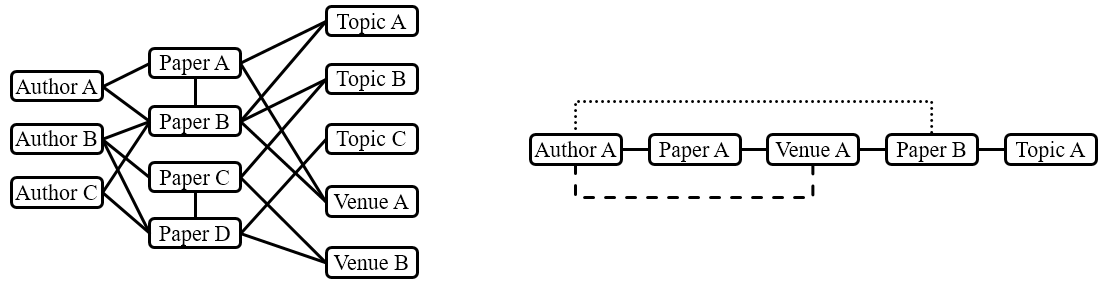}
	\caption{Citation network (left) and a random walk sampled from citation network (right). Solid line refers to explicit relation. Dashed line and dotted line refer to implicit relations using one intermediate node and two intermediate nodes, respectively.}
	\label{fig:citation-network}
\end{figure}

\section{Introduction}

Network embedding has been actively researched in the data mining field~\cite{perozzi2014deepwalk, tang2015line, grover2016node2vec, dong2017metapath2vec, fu2017hin2vec, hussein2018meta, shi2018aspem}. To apply machine learning techniques such as classification and regression \cite{murphy2012machine} to network data, nodes in a graph are generally mapped to low-dimensional embedding vectors. These vectors are fed into various downstream tasks such as node classification, link prediction and visualization \cite{dong2017metapath2vec}. As a heterogeneous network can represent nodes and edges of different types within a single network, it has been widely used to represent diverse underlying information in real-world data. Citation network (Fig. \ref{fig:citation-network}) is a representative heterogeneous network, which typically consists of four types of nodes -- author, paper, topic and venue \cite{dong2017metapath2vec, fu2017hin2vec, shi2018aspem, sun2011pathsim, yu2012citation}. These nodes are connected using relation types (i.e. edges in the network) such as authorship, citation, related topic, and published venue, and the size of each relation type (i.e. the number of edges regarding the type) is different.

Random walk, which is to sample a sequence of nodes from a network, has been widely adopted as a basic tool for extracting information from a network \cite{perozzi2014deepwalk,grover2016node2vec}. Rooted at a node, the next node is repeatedly selected at random among its neighboring nodes, until a walk reaches a predefined length. Many network embedding methods \cite{perozzi2014deepwalk,grover2016node2vec} use random walk to sample a sequence from a network because it preserves explicit and implicit relations inside the network. Explicit relation refers to a direct relationship between two nodes, and implicit relation refers to an indirect relationship between two nodes connected via some intermediate nodes. For example, in the random walk from the citation network (Fig. \ref{fig:citation-network}), the relationship between $Author A$ and $Paper A$ is explicit, whereas that between $Author A$ and $Venue A$ is implicit, as they are connected via $Paper A$.

However, random walk produces an imbalanced training of heterogeneous networks, because the major relation types take a large portion of training samples and thus dominate the training and minor relation types will hardly be learned. For example, in YAGO knowledge base, the number of explicit relations between $Person$ and $Position$ is 2,012, but that of explicit relations between $Person$ and $Organization$ is 886,529 (Refer to Table \ref{tab:dataset_statistic}). Thus, the $Person-Position$ relationship is unlikely to be captured by existing random walk strategies, as it is dominated by the $Person-Organization$ relationship. Simply incorporating an equal number of samples for each relation type will not suffice either, because the learning difficulty is different for each relation type. Instead, we regulate the number of samples in a sampled random walk by considering the loss value which represents the learning difficulty. Recently proposed heterogeneous network embedding methods \cite{fu2017hin2vec,hussein2018meta} do not handle the imbalance problem, because they just use the conventional random walk or fix the ratio, by a hyper-parameter, which determines whether to switch the relation type when sampling the next node in a random walk.

\begin{figure}
	\centering
	\includegraphics[width=\linewidth]{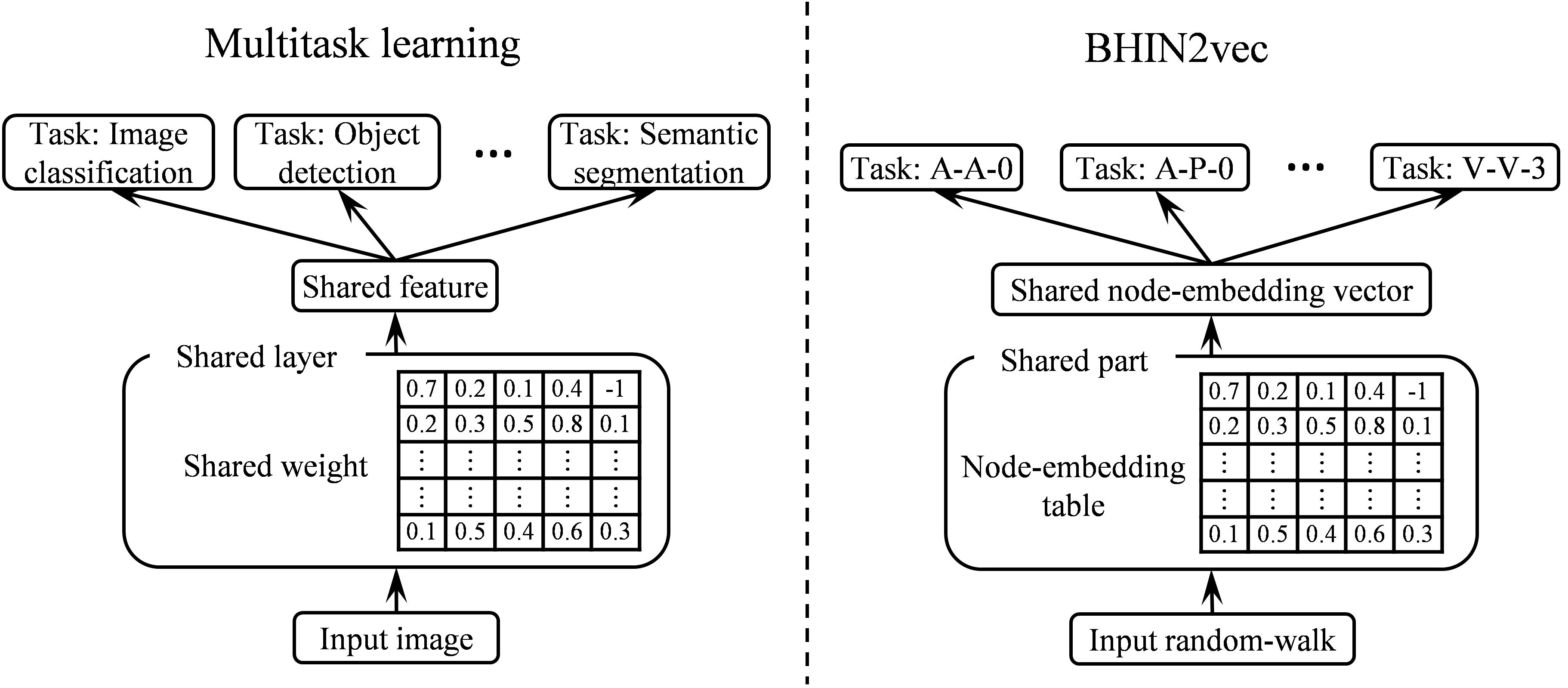}
	\caption{Overall diagram of the multitask learning model and BHIN2vec}
	\label{fig:conceptual-diagram}
	\vspace{-0.2cm}
\end{figure}

In this paper, we propose \textit{Balanced Heterogeneous Information Network to Vector}, called BHIN2vec, to resolve the imbalance issue in the heterogeneous network. The core idea of BHIN2vec is to view the heterogeneous network embedding as a multi-task learning problem in that each task is derived from a relation type in a network. For example, the goal of a task "$Author-Author-0$" is to maximize the similarity of an explicit relation between two authors (Fig. \ref{fig:conceptual-diagram}). We adopt GradNorm \cite{chen2018gradnorm}, which is a technique that employs inverse training ratio to solve the training imbalance issue in multi-task learning, to balance the training of different relation types in a network so as to generate a balanced embedding vector for each node. More precisely, we use inverse training ratio to first find the relation types that are less trained, i.e., relations that incur high loss values, and then generate random walks that contain more of the less-trained relations. In doing so, we propose a new random walk strategy that introduces a stochastic matrix to store the probabilities of choosing the next node type given the current one. The stochastic matrix is trained using inverse training ratio, and it eventually generates a random walk that contains more of the less-trained relations, and that considers all possible relation types.

BHIN2vec has three benefits compared to other network embedding methods. First, BHIN2vec trains all different relationships in a balanced way, and thus facilitates high quality representations for minor relations. Also, in the experimental section, we demonstrate that containing the minor relations in the embedding space actually helps represent the major relations. Second, BHIN2vec considers all possible relation types by introducing a stochastic matrix, which is a more elaborate way to handle type information within a heterogeneous network. Finally, BHIN2vec produces the inverse training ratio for each relation. This statistic can serve as a measure of how much each relation is mapped to the embedding space. By visualizing the stochastic matrix that stores the information of inverse training ratio, we can understand which type of relation is reflected well. To the best of our knowledge, this is the first work to leverage a new statistic apart from the training loss for heterogeneous network embedding.

The main contributions of our paper are summarized as follows:
\begin{itemize}
	\item We propose a novel heterogeneous network embedding method that automatically handles the imbalance issue in a heterogeneous network.
	\item We provide the stochastic matrix trained from inverse training ratio to understand how much each relation is represented compared to other relations during training.
	\item We conduct node classification and recommendation task to demonstrate that our embedding vector outperforms the state-of-the-art methods.
\end{itemize}
The rest of this paper is organized as follow. In section 2 and 3, we explain the related work and define our problem and widely used notations. In section 4, we propose our method, BHIN2vec. In section 5, we report our experimental results and discuss them. Finally, we conclude our research in section 6.

\section{Related work}

\subsection{Heterogeneous network embedding}
The goal of network embedding is to transform nodes in a network into a embedding space while preserving their property. Two approaches exist for the network embedding: factorization based methods \cite{zhang2018arbitrary} and random walk based methods \cite{perozzi2014deepwalk,grover2016node2vec}. We focus on the random walk based methods. The random walk based methods sample random walks from a network and maximize the similarity between two nodes contained in the random walks \cite{perozzi2014deepwalk,grover2016node2vec}. Deepwalk \cite{perozzi2014deepwalk} samples random walks without any constraint and other methods change the random walk strategy to capture the underlying structure in a network \cite{grover2016node2vec}.

Recent heterogeneous network embedding methods consider the type in a heterogeneous network \cite{fu2017hin2vec,dong2017metapath2vec,hussein2018meta}. One approach is to use a meta-path analyzed by domain experts. Metapath2vec \cite{dong2017metapath2vec} samples random walks controlled by the meta-path. However, finding an appropriate meta-path is hard as the number of types increases. Therefore, methods that do not require a meaningful meta-path have emerged \cite{fu2017hin2vec,hussein2018meta}. HIN2vec \cite{fu2017hin2vec} introduces a meta-path embedding table which saves the embedding vector for all possible meta-paths. These embedding vectors provide a different intensity for each dimension based on the corresponding meta-path. JUST \cite{hussein2018meta} designs a new random walk strategy without meta-path. They regulate the transition between types in a random walk using hyper-parameters.

\subsection{Balance training for multi-task learning}
Multi-task learning is to learn a single model that handles multiple tasks \cite{caruana1997multitask}. Multi-task models generate robust intermediate features because these features need to fit various tasks \cite{zhang2017overview,chen2018gradnorm,chen2011integrating}. These features are fed into the task-specific layers to perform tasks. Task-specific losses are calculated from the tasks and model optimizes an aggregated loss to jointly learn them. In the recent success of deep learning, computer vision is the primary application area for multi-task learning, but it can be used in various fields such as natural language processing \cite{sogaard2016deep} and speech synthesis \cite{wu2015deep}.

The balance among tasks is important when training a multi-task model \cite{kendall2018multi,chen2018gradnorm}. The characteristics for the tasks are diverse, which impedes the model to learn them compatibly. The irregular size of gradients can be one possible characteristic to inhibit training. The gradient size might be affected by the loss value and some large gradients can dominate the shared features. To resolve the imbalance, GradNorm \cite{chen2018gradnorm} introduces inverse training ratio which represents how much training goes along and adjusts the gradient size proportional to the training ratio.

\section{Problem Definition}
We formally define three concepts that are widely used in this field: heterogeneous network, meta-network and network embedding.

\paragraph{Heterogeneous network}
A heterogeneous network consists of four components, $G = \left( V, E, T, \phi \right)$. $V = \left\{ v_i \; | \; i \in \mathbb{N} \right\}$ and $T = \left\{ t_i \; | \; i \in \mathbb{N} \right\}$ refer to the node set and their type, respectively. $E = \left\{ \left( v_i, v_j \right) \; | \; v_i \in V \land v_j \in V \right\}$ refers to the edge set. The mapping function, $\phi \left( \cdot \right) : V \rightarrow T$, indicates the type for each node.

\paragraph{Meta-network} Given a heterogeneous network $G = \left( V, E, T, \phi \right)$, a meta-network is defined as $G_{meta} = \left( V_{meta}, E_{meta} \right)$. This network represents the relationship between node types in the heterogeneous network. The node for meta-network is the node type in the heterogeneous network, $V_{meta} = T$. Two node types are connected if there exist at least one corresponding connection in the network, $E_{meta} = \left\{ (t_i, t_j) \; | \; ^\exists \left(v_x, v_y\right) \in E \land \phi\left( v_x \right) = t_i \land \phi\left( v_y \right) = t_j \right\}$.

\paragraph{Network embedding} Given a network $ G $, the network embedding is to find a function $ f \left( \cdot \right) : V \rightarrow \mathbb{R}^{d} $ which takes one node as input and gives an embedding vector for that node as output. The embedding vector captures the network structure.

\section{Method}

\begin{figure*}
	\centering
	\includegraphics[width=\linewidth]{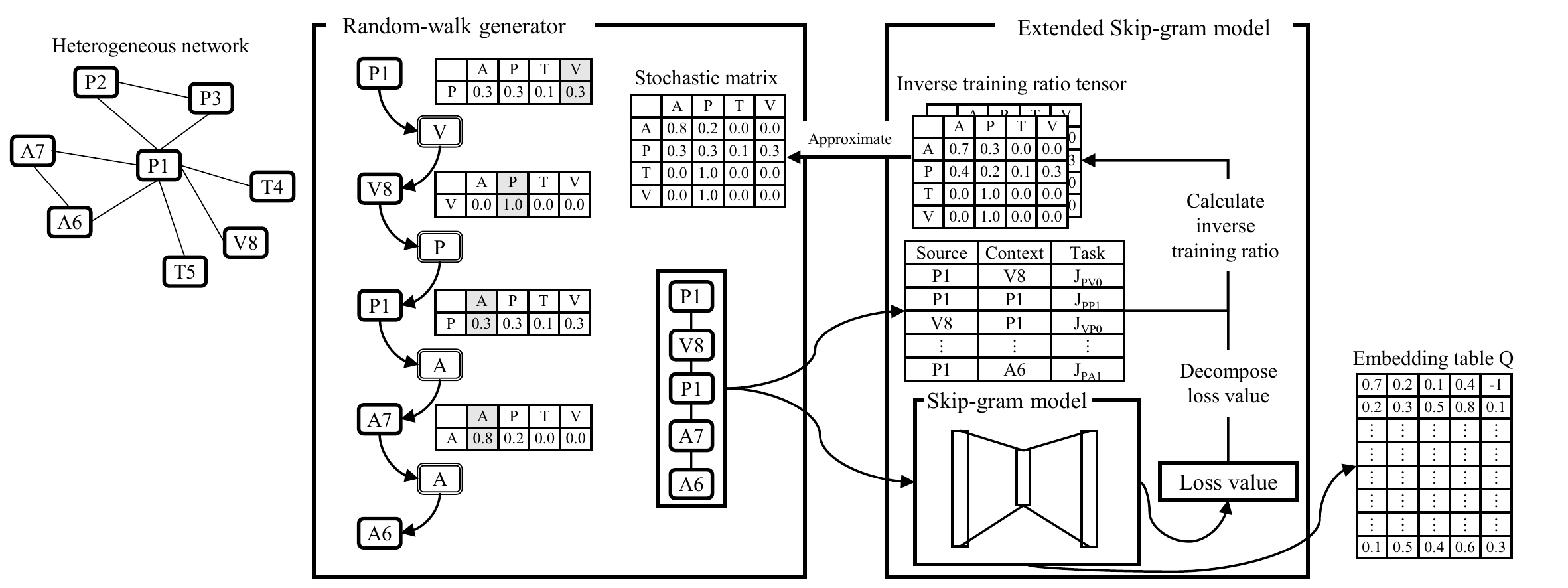}
	\caption{The conceptual diagram of BHIN2vec. We use a simple citation network. The node name consists of the node type and their index. The walk length is 5, the context window size is 2, the node dimension is 5. The random walk generator uses the corresponding row in the stochastic matrix to sample next type. In the extended skip-gram model, we split loss value based on the relation type. The inverse training ratio tensor is produced as a side-product and used for training the stochastic matrix in the random walk generator.}
	\label{fig:bhin2vec_framework}
\end{figure*}

Our proposed method, BHIN2vec, consists of two parts: a skip-gram model and a biased random walk generator. We extend the skip-gram model and create a side-product, called inverse training ratio tensor. In doing so, we propose a new random walk strategy that uses a stochastic matrix. We connect our extended skip-gram model and the new random walk strategy by training the stochastic matrix using the inverse training ratio tensor. The conceptual diagram of BHIN2vec for citation network is illustrated in Figure \ref{fig:bhin2vec_framework}.

\subsection{BHIN2vec: Skip-gram model}

\paragraph{Normal skip-gram model}
The skip-gram model has a node embedding table $ Q \in \mathbb{R}^{\left| V \right| \times d } $ which stores the $d$-dimensional embedding vector for all nodes \cite{mikolov2013distributed,perozzi2014deepwalk}. Our objective is to learn $ Q $ where $ f(v_i) = Q[i] $. The model takes a sequence of nodes $ w = \left( w_1, \cdots, w_l \right) $ with length $ l $, called walk, as an input. The skip-gram model maximizes the probability to predict context nodes using a source node. Choosing a source node $v_i$ for a given walk, the context nodes for the $v_i$ are the $k$ nodes right behind the $v_i$ in the walk. $k$ is a predefined parameter, called the context window size. We calculate the inner product of the embedding vector for the source node and the embedding vector for the context node and apply the softmax function to get the probability to predict the context node given the source node. Using the negative log-likelihood loss, the skip-gram loss for one random walk $w$ is calculated as
\begin{equation}
\resizebox{.91\linewidth}{!}{$
	\displaystyle
	L = - \sum_{i=1}^{l} \sum_{j=1}^{k} \log p \left( w_{i+j} \, | \, w_i \right) = - \sum_{i=1}^{l} \sum_{j=1}^{k} \log \frac{e^{f \left( w_{i+j} \right)^\top f \left( w_i \right)}}{\sum_{v_n}^{V} e^{f \left( v_n \right)^\top f \left( w_i \right)}} \text{.}
	$} \label{basic_skipgram}
\end{equation}
The denominator of the softmax function requires large computations, so we adopt the negative sampling approach to approximate that value \cite{mikolov2013distributed}. We take $m$ samples, called $N_V$, from $V$ with their empirical distribution. Then, the above loss function is changed to 
\begin{gather}
	L = - \sum_{i=1}^{l} \sum_{j=1}^{k} \left( L_{p} \left(w_{i+j}, w_{i} \right) + \sum_{v_o}^{N_V} L_{n} \left( v_o, w_i \right) \right) \label{loss_negative_sampling}\\ 
	L_p \left( v_c, v_s \right) = \log \sigma \left( f \left( v_c \right)^\top f \left( v_s \right) \right) \label{eq:Lp}\\
	L_n \left( v_c, v_s \right) = \log \sigma \left( - f \left( v_c \right)^\top f \left( v_s \right) \right) \label{eq:Ln} \text{,}
\end{gather}%
where $ \sigma $ is the sigmoid function. We calculate the gradient of loss value with respect to the embedding table and update the embedding table to minimize the loss value.

\paragraph{Multi-task setting for heterogeneous network embedding}
To apply the technique used in multi-task learning, we reconstruct heterogeneous network embedding as if we train multiple tasks simultaneously. First, we define virtual tasks using the relation type. The definition of virtual task $J_{ijk}$ is to predict nodes which types are $t_j$ given a node which type is $t_i$ and the two nodes are connected via $k$ intermediate nodes. Then, we construct a possible task set that contains the corresponding relation type which is used in the skip-gram model. Given a walk $w$, the skip-gram model uses the relations between two nodes where two nodes are connected via at most $k$ nodes. Considering all combination of the source node's type and context node's type, the number of tasks that are used in the skip-gram model is $k \times \left|T\right| \times \left|T\right|$ where $k$ is the context window size and $\left|T\right|$ is the number of types in a network. We reduce the size of possible tasks by removing the tasks that always not appear in any random walks. If an implicit relation between $t_i$ and $t_j$ via $k$ nodes doesn't exist in meta-network, then the task $J_{ijk}$ is always not contained in any walk. Finally, we define a possible task set $\mathit{J}_{possible}$ which can be contained in a random walk.
\begin{gather}
	\mathit{J}_{possible} = \left\{ J_{xyz} | \left(A^{z}\right)_{xy} > 0 \right\}\\
	A = \text{the adjacency matrix of } G_{meta}
\end{gather}

\paragraph{Balance in multitasks}
To quantify the imbalance in virtual tasks, we formulate the loss value for each task and define inverse training ratio tensor. We split the $L_p$ and $L_n$ in Equation (\ref{eq:Lp}) and (\ref{eq:Ln}) for each virtual task and calculate the loss values.
\begin{gather}
\resizebox{.91\linewidth}{!}{$
	\displaystyle
	L \left[ J_{xyz} \right] = - \frac{\sum_{i=1}^{l} \left( L \left[ J_{xyz} \right]_{p} \left(w_{i+z+1}, w_{i} \right) + \sum_{v_o}^{N_V} L\left[ J_{xyz} \right]_{n} \left( v_o, w_i \right) \right)}{\sum_{i=1}^{l} \left( \mathbb{I} \left[ J_{xyz} \right] \left(w_{i+z+1}, w_{i} \right) + \sum_{v_o}^{N_V} \mathbb{I} \left[ J_{xyz} \right] \left( v_o, w_i \right) \right)} \label{loss_per_task}
	$}\\
	\resizebox{.91\linewidth}{!}{$
		\displaystyle
	L \left[ J_{xyz} \right]_p \left( v_c, v_s \right) = 
	\begin{cases}
	L_p \left( v_c, v_s \right) & \text{if } \phi \left( v_c \right) = t_y \land \phi \left( v_s \right) = t_x \\
	0 & \text{otherwise}
	\end{cases}
	$}\\
	\resizebox{.91\linewidth}{!}{$
	\displaystyle
	L \left[ J_{xyz} \right]_n \left( v_c, v_s \right) = 
	\begin{cases}
	L_n \left( v_c, v_s \right) & \text{if } \phi \left( v_c \right) = t_y \land \phi \left( v_s \right) = t_x \\
	0 & \text{otherwise}
	\end{cases}
	$}\\
	\mathbb{I} \left[ J_{xyz} \right] \left( v_c, v_s \right) = 
	\begin{cases}
	1 & \text{if } \phi \left( v_c \right) = t_y \land \phi \left( v_s \right) = t_x \\
	0 & \text{otherwise}
	\end{cases}
\end{gather}
Also, some relation types are not contained in a random walk by chance. In this case, we take the previous loss calculated in the previous iteration as the surrogate value for the current one. For the loss value for each task, we take the mean to normalize the number of occurrence. These loss values represent which task doesn't fit in the embedding space compared to the others. Given the loss values, we define the inverse training ratios as
\begin{gather}
	\tilde{L} \left[ J_{xyz} \right] \left( t \right) = L \left[ J_{xyz} \right] / L_{initial} \left[ J_{xyz} \right]\\
	r \left[ J_{xyz} \right] \left( t \right) = \tilde{L} \left[ J_{xyz} \right] \left( t \right) / \mathbb{E}_{\mathit{J}_{possible}} \left[ \tilde{L} \left[ J \right] \left( t \right) \right] \label{inverse_training_rate} \text{,}
\end{gather}
where $L_{initial} \left[ J_{xyz} \right]$ is the initial loss when training starts \cite{chen2018gradnorm}.\footnote{Because the initial loss is unstable, we take a theoretical loss value as the initial loss. Suppose that $p \left( w_{i+j} | w_i \right) = 0.5$ in the Equation \eqref{basic_skipgram}, the initial loss value is $k \times (l-k) \times 0.6931$.} $\tilde{L} \left[ J_{xyz} \right] \left( t \right)$ is the training ratio of $J_{xyz}$ at time $t$, representing the amount of training that has been done.
$r \left[ J_{xyz} \right]$ is the relative inverse of the training ratio.
If $r \left[J_{xyz}\right] \left(t\right) > 1$, the task $J_{xyz}$ is premature compared to the other tasks.
We arrange these inverse training ratios to a 3-dimensional tensor $ I \in \mathbb{R}^{k \times \left| T \right| \times \left| T \right|} $, called inverse training ratio tensor.
We set one for the tasks that always not occur in a random walk.
\begin{equation}
	I_{zxy} \left( t \right) = \begin{cases}
		r [ J_{xyz} ] \left( t \right) & \text{if } J_{xyz} \in \mathit{J}_{possible} \\
		1 & \text{otherwise}
	\end{cases}
	\label{inverse_training_rate_tensor}
\end{equation}

\paragraph{Heterogeneous skip-gram model}
We adopt two variants for the skip-gram model. We sample negative nodes which have the same type with positive node \cite{dong2017metapath2vec}. It gives stable loss value for each task. Also, we create a task embedding table $Q_R \in \mathbb{R}^{k \times \left|T\right| \times \left|T\right| \times d}$ and define function $f_R \left( k, t_i, t_j \right) = Q_R \left[ k, t_i, t_j \right]$ which maps each task into $d$-dimensional vector. By multiplying the node embedding vector by the corresponding task embedding vector, relations will be embedded with different intensities for each dimension.
\begin{gather}
	L_p \left( v_c, v_s \right) = \log \sigma \left( \left( \sqrt{r} \odot f \left( v_c \right) \right)^\top \left( \sqrt{r} \odot f \left( v_s \right) \right) \right) \\
	L_n \left( v_c, v_s \right) = \log \sigma \left( - \left( \sqrt{r} \odot f \left( v_c \right) \right)^\top \left( \sqrt{r} \odot f \left( v_s \right) \right) \right) \\
    r = f_R \left( k, \phi \left( v_c \right), \phi \left( v_s \right) \right) 
\end{gather}

\subsection{BHIN2vec: Biased random walk generator}
\paragraph{Random walk strategy}
To provide a walk to the skip-gram model, we sample random walks from a network. Rooted at a node, we randomly select the next node from the adjacent nodes. Iterating this procedure $l$ times, a walk with length $l$ is generated from the network. To extract informative relations in the heterogeneous network, the random walk strategy needs to consider type. When sampling a random walk, a simple extension to consider type is to do an additional sampling for type. More precisely, we determine the type for the next node by sampling and do another sampling for the next node that has the sampled type. The remaining part of the random walk strategy is how to sample the next type and the next node.
 
\paragraph{Stochastic matrix}
We introduce a stochastic matrix to do bias sampling when sampling the next type. The stochastic matrix $P \in \mathbb{R}^{\left| T \right| \times \left| T \right|}$ describes the transition probabilities from the current states to the next state \cite{gagniuc2017markov}. In our case, the value $P_{ij} = p \left( t_j | t_i \right)$ in the stochastic matrix $P$ is the probability to choose $t_j$ for the next type when the current node type is $t_i$.
\begin{equation}
	P_{ij} = p \left( t_j | t_i \right) \; \text{such that} \; \sum_{j} P_{ij} = 1
\end{equation}
We set $p \left( t_j | t_i \right)$ to zero if no edge between $t_i$ and $t_j$ exists in the meta-network. So, this stochastic matrix can be viewed as a weighted adjacency matrix for the meta-network. We use the corresponding row of the current type in the stochastic matrix when sampling the next type. Also, note that the stochastic matrix represents the multi-hop type transition probability in a compact manner.
\begin{equation}
	(P^k)_{ij} = p \left( t_j | t_i, k \right)
\end{equation}
Based on this probability, we approximate the ratio for the implicit relation included in the sampled random walk. The pseudo-code for our biased random walk is illustrated in the Algorithm \ref{alg:algorithm}.
\begin{algorithm}[tb]
	\caption{BHIN2vecWalk}
	\label{alg:algorithm}
	\KwIn{Start node $ v $, stochastic matrix $ P $, Walk length $ l $}
	\myinput{Heterogeneous network $ G = \left( V, E, T, \phi \right)$}
	\KwOut{Random walk $ w $}
	$ w[1]=v $\;
	\For{$i\gets1$ \KwTo $l-1$}{
		$ nxt\_t = BiasedSample \left( T, P \left[ \phi \left( w[i] \right), : \right] \right) $\;
		$ target = \{ v_j \, | \, \left( w[i], v_j \right) \in E \land \phi \left( v_j \right) = nxt\_t \} $\;
		$ w[i+1] = UniformSample \left( target \right) $\;
	}
	\textbf{return} $ w $
\end{algorithm}

\subsection{BHIN2vec: From inverse training ratio tensor to stochastic matrix}
We train the stochastic matrix to store the information in the inverse training ratio tensor.
We intend to sample more of less-trained relations so that the less-trained relations would be reflected more in the embedding space.

\begin{algorithm}[tb]
	\caption{BHIN2vec}
	\label{alg:algorithm2}
	\KwIn{Heterogeneous network $ G = \left( V, E, T, \phi \right) $}
	\myinput{Latent dimension $ d $, walk length $ l $}
	\myinput{Context window size $ k $, negative sample size $ m $}
	\myinput{Epoch num $ e $, learning rate $ r, r_2 $}
	\KwOut{Node embedding matrix $ Q $}
	Initialize $ Q \in \mathbb{R}^{\left| V \right| \times d} $\;
	$ P = [[0, \cdots, 0], \cdots, [0, \cdots, 0]] \in \mathbb{R}^{\left| T \right| \times \left| T \right|} $\;
	$ P[i, j] = 1 $ for explicit edge\; 
	\For{$ 1 $ \KwTo $ e $}{
		\ForEach{$ v \in V $}{
			$ w = BHIN2vecWalk \left( v, P, l, G \right) $\;
			$ L = SkipGram \left( w, k, m, Q \right) $\;
			$ Q = Q - r \frac{\partial}{\partial Q} L $\;
			Create $ I \in \mathbb{R}^{k \times \left| T \right| \times \left| T \right|} $ using Equation \eqref{inverse_training_rate_tensor}\;
			Calculate $ L_{stochastic} $ using Equation (\ref{loss_stochastic2})\;
			$ P = P - r_2 \frac{\partial}{\partial P} L_{stochastic} $\;
		}
	}
	\textbf{return} $ Q $
\end{algorithm}

\paragraph{Perturbation approach}
We perturb already existing stochastic matrix using inverse training ratio tensor.
We use a uniform stochastic matrix $ P_{uni} \in \mathbb{R}^{\left| T \right| \times \left| T \right|} $ as an existing solution.
In the uniform stochastic matrix, the transition probability to target type is equal for each source type.
\begin{equation}
	P_{uni_{xy}} = 
	\begin{cases}
		\frac{1}{degree(t_x)} &\text{if} \; (t_x, t_y) \in E_{meta} \\
		0 &\text{otherwise}
	\end{cases}
\end{equation}
Using the uniform stochastic matrix, we calculate the probability to move from $t_i$ to $t_j$ in $k$ steps and perturb this probability using $I_{kij}$.
\begin{equation}
 	L_{stochastic} = \sum_{i=0}^{k-1} \left| P^{i+1} - \left( P_{uni}^{i+1} + \alpha \left(I_i - \mathbf{1} \right) \right) \right|_F^2 \label{loss_stochastic2}
\end{equation}
The perturbation parameter $ \alpha $ determines how much the perturbation is applied to the existing solution.

\paragraph{Update stochastic matrix}
We calculate the gradient of this loss value with respect to the stochastic matrix and update the stochastic matrix to minimize the loss value.
To preserve the property in the stochastic matrix, we only update nonzero values and clip the values between zero and one.
Then, we normalize the row in the stochastic matrix.  
 
In summary, given a random walk, the skip-gram model updates the embedding table and creates an inverse training ratio tensor as a side-product.
Then, using the inverse training ratio tensor, the loss value for the stochastic matrix is calculated and used for training.
After updating the stochastic matrix, we sample a new random walk.
We alternately optimize the embedding table and the stochastic matrix by iterating this process until the training converges.
The pseudo-code for the overall BHIN2vec procedure is illustrated in the Algorithm \ref{alg:algorithm2}.

\section{Experiments}

In this section, we create embedding vectors using BHIN2vec and analyze the result to understand the behavior of BHIN2vec. We focus on the following research questions.
\begin{itemize}
	\item \textbf{RQ1}: Does our method get better node representations than other baselines?
	\item \textbf{RQ2}: Do the representations contain all different relation types in the heterogeneous network? 
	\item \textbf{RQ3}: How do the hyper-parameters affect our method? 
\end{itemize}
To answer \textbf{RQ1} and \textbf{RQ2}, we conduct two general tasks: multi-class node classification and recommendation using link prediction. To answer \textbf{RQ3}, we conduct the sensitivity analysis on the perturbation hyper-parameter $\alpha$.
 
\subsection{Dataset and Baseline}

\paragraph{Dataset} We build up heterogeneous networks from the real-world data. All datasets can be accessed from the public websites.
\begin{itemize}
	\item {\bf BlogCatalog} is the social blog directory which manages the bloggers and their groups.\footnote{http://socialcomputing.asu.edu/datasets/BlogCatalog3} This dataset contains two types: user and group. Friendships between users and group membership exist in the network.
	\item {\bf Douban} is a movie platform website. We used preprocessed version of this dataset.\footnote{https://bit.ly/2CDFI9z} Movie, actor, director and user nodes compose this network and the nodes are connected with four relation types: an actor participates a movie; a director produces a movie; a user watches a movie; and a friendship between two users.
	\item {\bf DBLP} is a website that manages the research publication.\footnote{https://aminer.org/citation} We used V10 for our experiment \cite{sinha2015overview}. We build up a heterogeneous network with four type: author, paper, topic and venue. The topic is generated by splitting the title and abstract into words. Four relationships exist in the network: an author writes a paper; a paper references other paper; a paper contains a topic and a paper is published in a venue.
	\item {\bf YAGO} is an open source knowledge base. We used preprocessed version for this experiment \cite{shi2018easing}. Seven types exist in the network: person, piece of work, prize, position, event, organization and location. Eight relation types exist in the network.
\end{itemize}
We removed nodes which degree is less than 2 because these nodes contain inadequate information. We summarize the statistic for each dataset in Table \ref{tab:dataset_statistic}.

\begin{table}
	\centering
	\caption{Dataset statistic}
	\begin{tabular}{@{}ccccc@{}}
		\toprule
		Dataset & \multicolumn{2}{c}{Node} & \multicolumn{2}{c}{Edge} \\ \midrule
		BlogCatalog & User & 10312 & U-U & 267181 \\
		& Group & 39 & U-G & 11581 \\ \midrule
		Douban & User & 12392 & M-U & 658412 \\
		& Movie & 9322 & U-U & 3268 \\
		& Actor & 5765 & M-A & 20400 \\
		& Director & 2202 & M-D & 6646 \\ \midrule
		DBLP & Author & 133774 & A-P & 667340 \\
		& Paper & 230356 & P-P & 996495 \\
		& Topic & 119190 & P-T & 2463562 \\
		& Venue & 165 & P-V & 219856 \\ \midrule
		YAGO & PErson & 346838 & LO-LO & 350005 \\
		& LOcation & 134817 & OR-LO & 5776 \\
		& WOrk & 71852 & EV-LO & 9171 \\
		& ORganization & 18716 & PE-LO & 177646 \\
		& EVent & 5590 & PE-OR & 886529 \\
		& PRize & 1711 & PE-WO & 298632 \\
		& POsition & 301 & PE-PO & 2012 \\
		&  &  & PE-PR & 69249 \\\bottomrule
	\end{tabular}
	\label{tab:dataset_statistic}
\end{table}

\paragraph{Baseline Method}
We compare our method with other network embedding methods.
\begin{itemize}
	\item {\bf Deepwalk} \cite{perozzi2014deepwalk} does a pioneering work by introducing the skip-gram model into network embedding task. This model samples random walks from a network and increases the similarity between the source nodes and context nodes.
	\item {\bf LINE} \cite{tang2015line} considers first and second order proximity in a network. Instead of sampling random walks, they directly optimize the similarity between nodes.
	\item {\bf HIN2vec} \cite{fu2017hin2vec} introduces a meta-path embedding table and jointly optimizes the node embedding table and the meta-path embedding table. They combine the node embedding vector with the meta-path embedding vector to control the intensity of each dimension for different relation types.
	\item {\bf JUST} \cite{hussein2018meta} adjusts the random walk strategy by introducing two hyper-parameters. Instead of using the meta-path, they sample random walks with the type transition probability and a queue that manages previously sampled types. 
\end{itemize}
We use the author codes except JUST because JUST doesn't publish their code.

\paragraph{Hyper-parameter tuning}
To focus on the heterogeneous property in the network, we fix all parameters related to the homogeneous property. All methods require a walk length $l$, the number of epoch $e$, the node dimension $d$, the context window size $k$ and the number of negative nodes per positive node $m$. We fix $l=100, e=10, k=5, m=5, d=128$ for all methods. Heterogeneous network embedding methods introduce different hyper-parameters to handle the type. HIN2vec doesn't have additional hyper-parameters. JUST takes two hyper-parameters, $\alpha$ and the size of queue. We perform grid search on $\alpha = \left[ 0.25, 0.5, 0.75 \right]$ and $\text{the size of queue} = \left[ 1, 2, 3 \right]$. Our method takes two hyper-parameters, the perturbation parameter $\alpha$ and the learning rate for the stochastic matrix $r_2$. We perform grid search on $\alpha = \left[ 0.05, 0.1, 0.2 \right]$ and $r_2 = \left[ 0.25, 0.025, 0.0025 \right]$. 

\begin{figure}
\centering
\includegraphics[width=\linewidth]{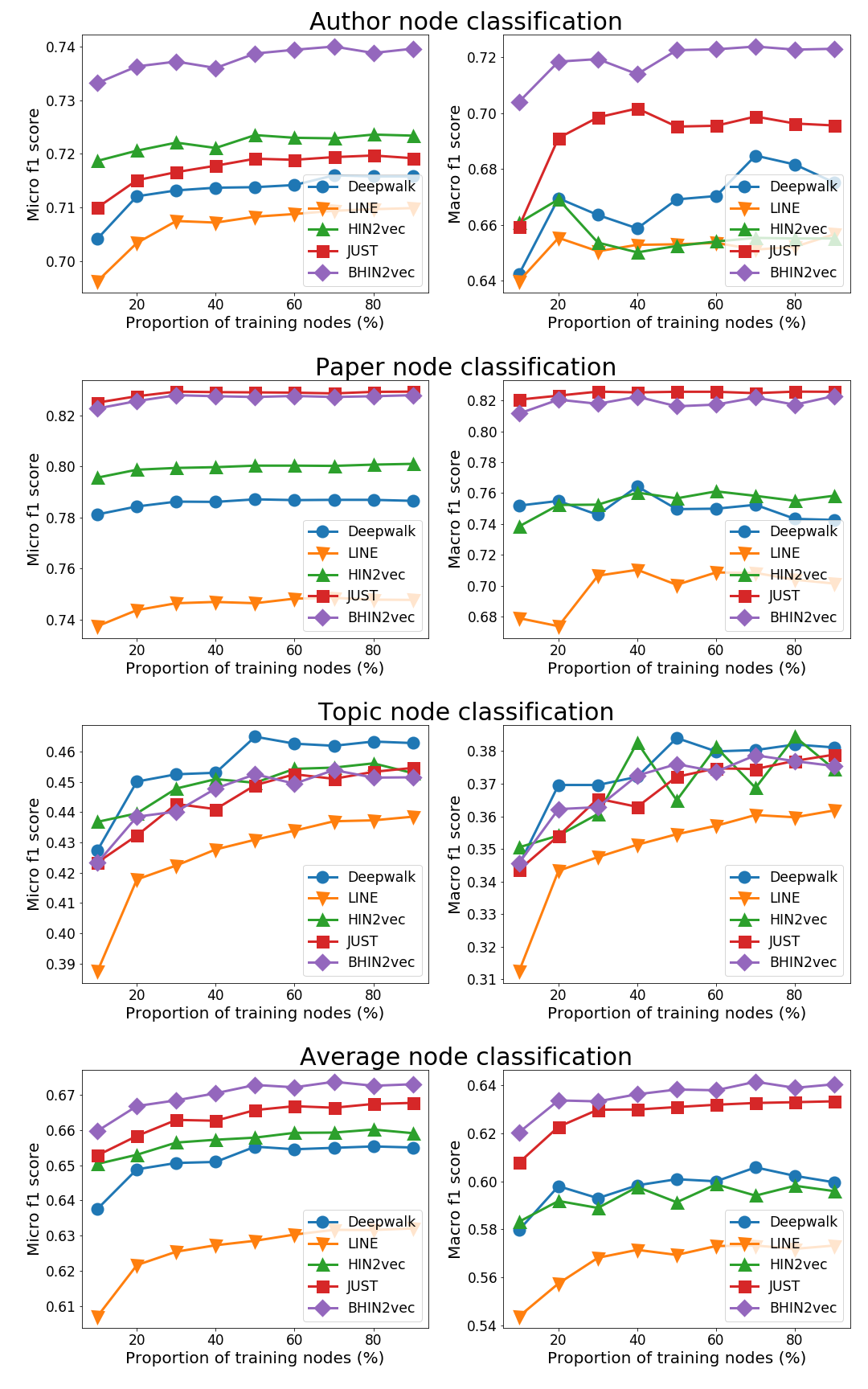}
\caption{Node classification result on DBLP dataset}
\label{fig:dblp_node_classification}
\end{figure}

\subsection{Multi-class node classification}
To answer \textbf{RQ1}, we measure the F1 score for the multi-class node classification task based on the node embedding table. Note that the purpose of this experiment is to evaluate the quality of the network embedding. We measure the performance of several node classification tasks and then determine the quality of network embedding based on their average. We conduct the multi-class node classification task on different node types.

\paragraph{Constructing label information}
We create a label for DBLP network. For venue node, We adopt the same way as Metapath2vec \cite{dong2017metapath2vec} does. In Google Scholar\footnote{https://scholar.google.com/citations}, we crawl top venues in each category and label venue nodes with their category. Then, we propagate the venue label to the paper nodes. We carefully determine the label for paper nodes which have diverse information. For a given paper, we collect the labels for papers citing that paper. The labels of the papers cited by that paper are also collected separately. Then, we choose the label that contains the most in collected labels. Finally, we randomly sample one of the three labels, the label found in each label set and the venue label, to determine the label of a given paper. The labels of author and topic nodes are determined by random sampling among the labels of connected papers.

\paragraph{Evaluation protocol}
We classify the nodes with given labels. First, we train a node embedding table using network embedding methods.
Then, we choose the embedding vectors for the labeled nodes. We randomly choose 20 percent of the labeled nodes to construct the test set and the remaining labeled nodes are used to construct the training set. We train the logistic regression classifier using the training set and evaluate the micro F1 and macro F1 score using the test set. Also, we conduct this experiment with different size of the training nodes. We report the average score from 10 repeated trials.

\paragraph{Result}
We reports the micro F1 and macro F1 values on $Author$, $Paper$, $Topic$ and its average in Figure \ref{fig:dblp_node_classification}. This experimental result has three noticeable points. First, our proposed method, BHIN2vec, gives better average micro F1 and macro F1 scores than the other methods do, which is the answer for \textbf{RQ1}. BHIN2vec reports comparable performance on $Topic$ nodes, and gives better micro F1 and macro F1 scores on $Author$ and $Paper$ nodes than that of the other methods. Second, BHIN2vec prevents some relationships from dominating the embedding space. In DBLP network, the $Paper-Topic$ relation is most common. Deepwalk \cite{perozzi2014deepwalk} and HIN2vec \cite{fu2017hin2vec} optimize that relation in the network. However, these methods does not optimize other minor relations inside the network, which results in the low F1 score in $Author$ and $Paper$ node classification task. Lastly, JUST \cite{hussein2018meta} achieves comparable F1 scores in $Topic$ nodes, but reports low F1 scores in $Author$ nodes because that method does not consider the $Paper-Topic$ relations and $Paper-Author$ relations separately. However, BHIN2vec distinguishes $Paper-Author$ relations and $Paper-Topic$ relations, which results in the improvement in $Author$ node classification. To understand how this mechanism works, we visualize the transition probability from paper to the other types in Figure \ref{fig:transition_probability}. In the early stage of training, BHIN2vec samples equal number of all possible relations in the random walks. As the $Paper-Topic$ relation becomes well reflected in the embedding space, the stochastic matrix is trained with this information and the transition probability for that relation decreases. Therefore, a random walk generated from BHIN2vec contains all possible relation type proportional to the loss value, which gives balanced embedding vectors.

\begin{table*}
	\centering
	\caption{Hit Rate @ 10 on Douban and BlogCatalog}
	\label{tab:Douban_recommendation}
	\begin{tabular}{@{}c|cccccccc|cccc@{}}
		\toprule
		Dataset & \multicolumn{8}{c|}{Douban} & \multicolumn{4}{c}{BlogCatalog} \\ \midrule
		Source node & \multicolumn{2}{c}{User} & \multicolumn{3}{c}{Movie} & Actor & Director & \multirow{2}{*}{Average} & \multicolumn{2}{c}{User} & Group & \multirow{2}{*}{Average} \\
		Target node & User & Movie & User & Actor & Director & Movie & Movie &  & User & Group & User &  \\ \midrule
		Deepwalk & 0.4630 & 0.2100 & 0.1200 & 0.4188 & 0.5607 & 0.3883 & 0.5060 & 0.3810 & 0.2079 & 0.3733 & \textbf{0.4414} & 0.3409 \\
		LINE & 0.2577 & 0.7142 & 0.5312 & 0.3672 & 0.4021 & 0.2694 & 0.2993 & 0.4059 & 0.4821 & 0.3734 & 0.2399 & 0.3651 \\
		HIN2vec & 0.3751 & 0.7790 & \textbf{0.6567} & 0.5374 & 0.5590 & 0.4169 & 0.4356 & 0.5371 & 0.4794 & 0.4260 & 0.4109 & 0.4388 \\
		JUST & 0.6045 & 0.4198 & 0.2664 & 0.5454 & 0.6365 & 0.5051 & 0.6107 & 0.5126 & 0.4631 & 0.3129 & 0.2992 & 0.3584 \\
		BHIN2vec & \textbf{0.6392} & \textbf{0.7925} & 0.6485 & \textbf{0.6277} & \textbf{0.7334} & \textbf{0.5865} & \textbf{0.7154} & \textbf{0.6776} & \textbf{0.6531} & \textbf{0.4314} & 0.3709 & \textbf{0.4851} \\ \bottomrule
	\end{tabular}
\end{table*}

\subsection{Recommendation using link prediction}

To answer \textbf{RQ2}, we conduct recommendation tasks for all relation types in various heterogeneous networks.

\paragraph{Evaluation protocol}
Given a heterogeneous network, we remove 20\% of total edges for the test set. With 80\% of remaining edges, called the training set, we train the node embedding table using several network embedding methods. We perform recommendation tasks for all relation types in a network. A recommendation task recommends a target node for a source node. For example, we recommend a director for a movie. We add negative examples to the training set by sampling same number of arbitrary node pairs for each relation from the network. Using the node pairs and edge, we create the edge embedding vectors by applying Hadamard function to two node vectors in the training set \cite{fu2017hin2vec}. We train a logistic regression classifier using the edge embedding vectors and evaluate the hit rate at 10 for the recommendation task, which is widely used in the recommendation field \cite{yin2017sptf, he2017neural, ebesu2018collaborative}. Given an edge $(v_i, v_j)$ in test set, we sample 99 nodes in which the sampled node type is same as the type of $v_j$. Note that the node pair created with $v_i$ doesn't occur in both the training set and the test set. Given 99 pairs of nodes and 1 edge, we create 100 edge embedding vectors and rank them using the trained classifier. If the edge is in the top-10 ranking list, we regard this result as successfully recommending the target node and increase the hit count. We apply this process to all edges in the test set and we report the proportion of hit. We report the average score from 5 repeated trials.

\begin{table}
	\setlength\tabcolsep{3pt}
	\centering
	\caption{Hit rate @ 10 on YAGO}
	\label{tab:yago_recommendation}
	\begin{tabular}{@{}ccccccc@{}}
		\toprule
		Source & Target & Deepwalk & LINE & HIN2vec & JUST & BHIN2vec \\ \midrule
		OR & LO & 0.4874 & 0.9253 & 0.9732 & 0.8914 & \textbf{0.9770} \\
		LO & OR & 0.8620 & 0.3290 & \textbf{0.8808} & 0.7311 & 0.8196 \\
		EV & LO & 0.8491 & 0.9612 & 0.9698 & 0.9612 & \textbf{0.9819} \\
		LO & EV & 0.7935 & 0.7027 & 0.8912 & 0.6649 & \textbf{0.8973} \\
		LO & LO & 0.8866 & 0.7961 & 0.9637 & 0.9306 & \textbf{0.9725} \\
		PE & OR & 0.9453 & 0.5261 & 0.9917 & 0.9832 & \textbf{0.9936} \\
		OR & PE & 0.9786 & 0.3244 & \textbf{0.9869} & 0.9715 & 0.9861 \\
		PE & PR & 0.8103 & 0.7751 & 0.9503 & 0.9445 & \textbf{0.9654} \\
		PR & PE & 0.9094 & 0.5611 & 0.8795 & 0.9113 & \textbf{0.9232} \\
		PE & WO & 0.7524 & 0.3262 & \textbf{0.8922} & 0.6490 & 0.8404 \\
		WO & PE & 0.9293 & 0.8367 & \textbf{0.9760} & 0.7903 & 0.9370 \\
		PE & PO & 0.6667 & 0.6667 & 0.6290 & 0.7333 & \textbf{0.7351} \\
		PO & PE & 0.8409 & 0.6250 & 0.8364 & 0.8409 & \textbf{0.8886} \\
		PE & LO & 0.5175 & 0.8462 & \textbf{0.9315} & 0.7679 & 0.9116 \\
		LO & PE & 0.6653 & 0.2257 & \textbf{0.6693} & 0.4601 & 0.6603 \\
		\multicolumn{2}{c}{Average} & 0.7930 & 0.6285 & 0.8948 & 0.8154 & \textbf{0.8993} \\ \bottomrule
	\end{tabular}
\end{table}

\begin{figure*}
	\centering
	\includegraphics[width=\linewidth]{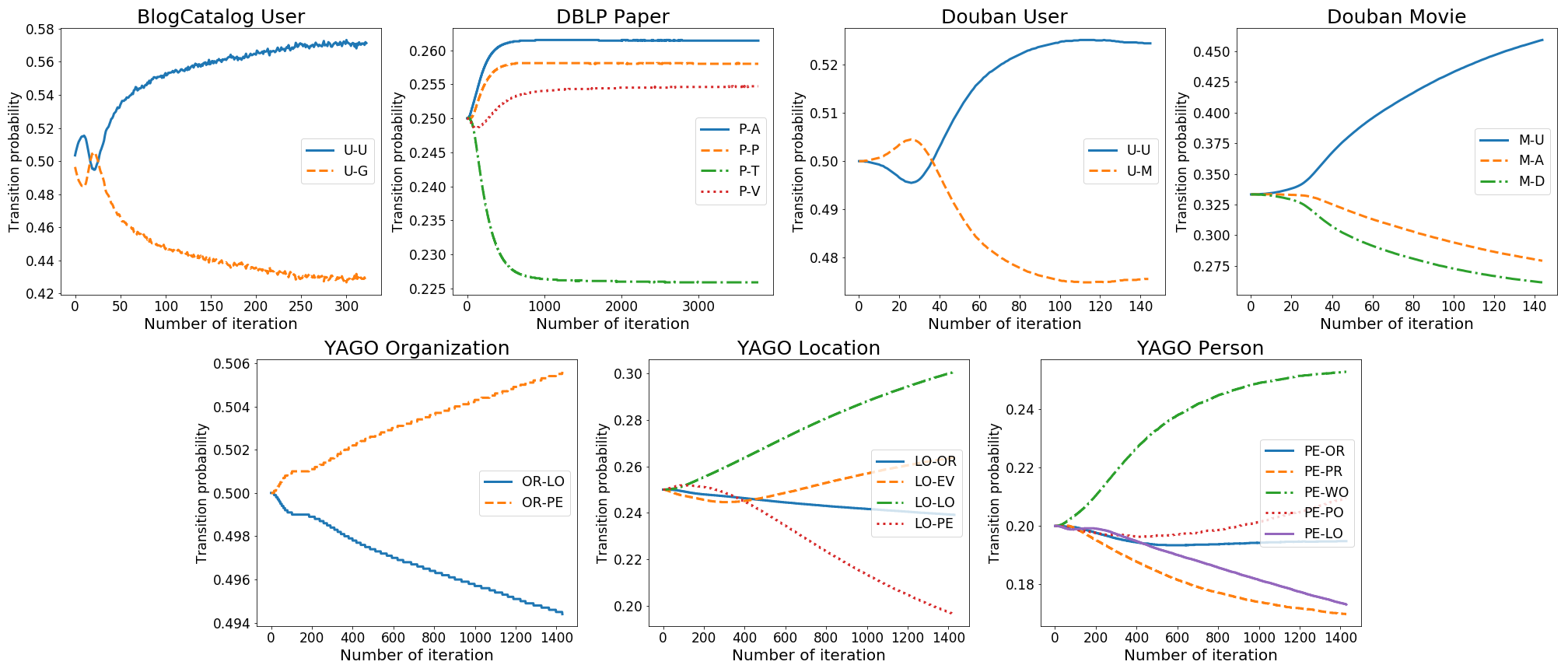}
	\caption{Transition probabilities in the stochastic matrix. We report the transition probabilities that give the highest average micro F1 score for DBLP and the highest average hit rate score for the other datasets.}
	\label{fig:transition_probability}
\end{figure*}

\paragraph{Results} 
The hit rate at 10 on Douban and BlogCatalog dataset is reported in the Table \ref{tab:Douban_recommendation}. We report the average score of all tasks to measure the overall quality of the embedding vectors. BHIN2vec gives high hit rates at 10 in overall relations, whereas other methods cannot learn all relation types in the heterogeneous network evenly. This result empirically shows that BHIN2vec contains all different relation types in one embedding vector, which is the answer for \textbf{RQ2}. Other methods focus on a specific relation. In the Douban network, Deepwalk \cite{perozzi2014deepwalk} focuses on the $Movie-Director$ relation and LINE \cite{tang2015line} focuses on the $User-Movie$ relation. HIN2vec \cite{fu2017hin2vec} represents the $User-Movie$ relations well because those relations are contained mostly in that network. BHIN2vec learns not only the $User-Movie$ relations but also other minor relations well, which acts as side information for embedding the $User-Movie$ relation in the embedding space. Therefore, the hit rate for that relation is better than HIN2vec does. JUST \cite{hussein2018meta} reports low hit rate in the $User-Movie$ relations. We conclude that JUST \cite{hussein2018meta} uses uniform probability when jumping to other types, so random walks cannot contain enough number of $User-Movie$ relations to train. BHIN2vec considers the loss value for all possible relation types and focuses on $User-Movie$ relations which report high loss value. We visualize the transition probability from $Movie$ type to other types (Fig. \ref{fig:transition_probability}). The probability from $Movie$ to $User$ type increases, which means that $User-Movie$ relations report a high loss value compared to other relations. So, BHIN2vec reports high hit rate score by including enough $User-Movie$ relations in the random walks to train.

In BlogCatalog, BHIN2vec improves both the $User-User$ task and the $User-Group$ task in a balanced way. However, BHIN2vec reports low hit rate for $Group-User$ task because if the $User-User$ relations are well represented in the embedding space, the group can not distinguish between two connected users. Therefore, $User-User$ relations and $User-Group$ relations show a tradeoff. We conclude that BHIN2vec finds the equilibrium point in this situation.

\begin{figure}
	\centering
	\begin{subfigure}[b]{0.475\textwidth}
		\includegraphics[width=\linewidth]{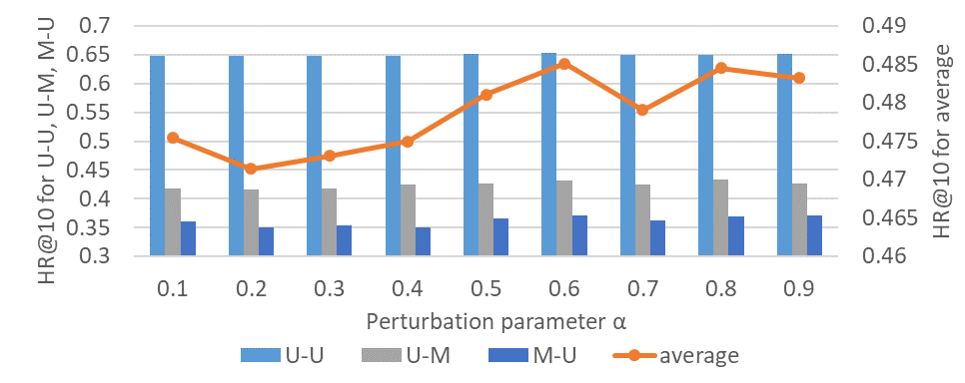}
		\caption{BlogCatalog dataset}
	\end{subfigure}
	\begin{subfigure}[b]{0.475\textwidth}
		\includegraphics[width=\linewidth]{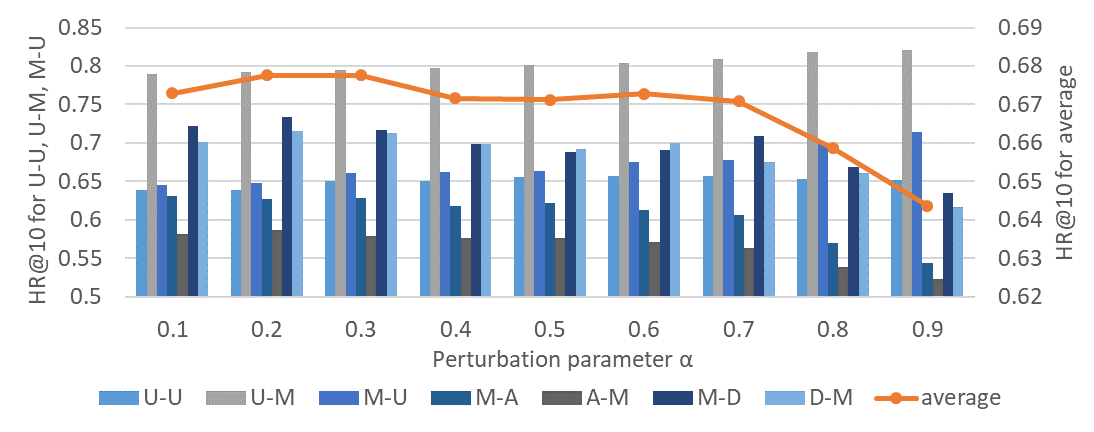}
		\caption{Douban dataset}
	\end{subfigure}
	\caption{Hit rate scores with varying perturbation hyper-parameter $\alpha$. Each bar represents a recommendation task. The source node type and target node type for a task are connected with a hyphen.}
	\label{fig:sensitivity-analysis}
\end{figure}

To scale up network size, we report the hit rate on YAGO network in the Table \ref{tab:yago_recommendation}. BHIN2vec gives the better average hit rate than the other methods. Note that BHIN2vec embeds both the $Person-Position$ relations, which is representative major relation type, and the $Person-Organization$ relations, which is representative minor relation type, well in the same embedding space. It means that we resolve the imbalance issue in the heterogeneous network. However, the hit rates are dropped for some tasks. The $Person-Work$ relations and the $Person-Location$ relations are much more complicated than the other relations because the number of $Work$ and $Location$ nodes are larger than the number of $Organization$ nodes. Therefore, to handle the network which has many relation types with extremely different complexities in a balanced way will be our future work.

\paragraph{Sensitivity Analysis}
To answer \textbf{RQ3}, we investigate the impact of perturbation parameter $\alpha$ on recommendation task. In the Figure \ref{fig:sensitivity-analysis}, we report the hit rate with different perturbation parameter $ \alpha $. In the BlogCatalog network, the average hit rate at 10 increases as the $\alpha$ increases. So, instead of using a uniform type transition probability, creating a random walk that contains the relation types proportional to the task loss is more effective to embed a network. In the Douban network, we observe that the major relations, i.e. $User-Movie$ relations, increase as $\alpha$ increases. In the Figure \ref{fig:transition_probability}, a random walk contains the $User-Movie$ relations more. This trend is intensified as $\alpha$ increases, which results in the improvement of corresponding recommendation tasks. However, we also observe that the hit rate at 10 for minor relations decreases as $\alpha$ increases because $User-Movie$ relation breaks the similarity between other minor relations. In summary, as $\alpha$ increases, BHIN2vec creates embedding vectors that contains overall relationships in a balanced way, but for large networks, the tradeoff between some relation types exists so that we recommend to use small $\alpha$ to protect the minor relation types.

\section{Conclusion}
We observe the imbalance issue in the heterogeneous network and resolve this issue by designing a new heterogeneous network embedding method. To balance in all possible relation types, we focus on the relation types that are less trained in the embedding space. To quantify how much each relation type is trained, we introduce the idea in multi-task learning. We define virtual tasks in that each task represents each relation type in the heterogeneous network. Then, we calculate the loss values for each virtual task by splitting the skip-gram loss and compute the inverse training ratios which represent how much each relation type is embedded. To focus on the tasks which report high loss value, we propose a new random-walk strategy that samples a random walk that contains more of less-trained relations. For the compact representation, we introduce the stochastic matrix in the random-walk strategy and train that stochastic matrix to store the information in the inverse training ratio.

We demonstrate that BHIN2vec produces node embeddings that contains all possible relation types evenly. We use our node embeddings to conduct two general tasks: node classification and recommendation. In node classification, we evaluate the micro F1 and macro F1 score for all node types. Our node embeddings give better F1 scores for all node types. Especially, our node embeddings give better F1 scores in $Author$ node classification, which addresses the importance of considering all possible relation types. Also, visualizing the stochastic matrix, we understand the mechanism of BHIN2vec. In recommendation, we evaluate the hit rate at 10 in three different heterogeneous networks. BHIN2vec improves the hit rate at 10 in overall relation types in three different networks. In YAGO, BHIN2vec successfully embeds both the major relation type, i.e. $Person-Organization$ relations, and minor relation type, i.e. $Person-Position$ relations, at the same time. Also, we observe the tradeoff between complicated relation types, which will be our future work.

%
\begin{acks}
This research was supported by Basic Science Research Program through the National Research Foundation of Korea (NRF) funded by the Ministry of Science and ICT (MSIT) (No. 2016R1E1A1A01942642)
\end{acks}

%
\bibliographystyle{ACM-Reference-Format}
\balance
\bibliography{CIKM-base}

\end{document}